\documentclass[aps,prb,superscriptaddress,reprint,showpacs,amssymb,amsmath,floatfix,citeautoscript,longbibliography]{revtex4-2}

\usepackage{graphicx}
\usepackage{color}
\usepackage{amsmath,hyperref}
\usepackage{float,booktabs}
\usepackage{babel}
\usepackage{ulem}
\usepackage{braket}
\usepackage{array}

\newcolumntype{C}[1]{>{\centering\let\newline\\\arraybackslash\hspace{0pt}}m{#1}}

\begin{document}

\title{Spectroscopic evidence of Kondo-induced quasi-quartet in CeRh$_2$As$_2$}

\author{Denise~S.~Christovam}
       \affiliation{Max Planck Institute for Chemical Physics of Solids, N{\"o}thnitzer Str. 40, 01187 Dresden, Germany}
\author{Miguel Ferreira-Carvalho}
    \affiliation{Max Planck Institute for Chemical Physics of Solids, N{\"o}thnitzer Str. 40, 01187 Dresden, Germany}
    \affiliation{Institute of Physics II, University of Cologne, Z\"{u}lpicher Str. 77, 50937 Cologne, Germany}
\author{Andrea~Marino}
    \affiliation{Max Planck Institute for Chemical Physics of Solids, N{\"o}thnitzer Str. 40, 01187 Dresden, Germany}
\author{Martin~Sundermann}
        \affiliation{PETRA III, Deutsches Elektronen-Synchrotron DESY, Notkestraße 85, 22607 Hamburg, Germany}
       \affiliation{Max Planck Institute for Chemical Physics of Solids, N{\"o}thnitzer Str. 40, 01187 Dresden, Germany}
\author{Daisuke~Takegami}
       \affiliation{Max Planck Institute for Chemical Physics of Solids, N{\"o}thnitzer Str. 40, 01187 Dresden, Germany}
\author{Anna Melendez-Sans}
       \affiliation{Max Planck Institute for Chemical Physics of Solids, N{\"o}thnitzer Str. 40, 01187 Dresden, Germany}
\author{Ku Ding Tsuei}
    \affiliation{National Synchrotron Radiation Research Center, 101 Hsin-Ann Road, Hsinchu 30077, Taiwan}
\author{Zhiwei Hu}
    \affiliation{Max Planck Institute for Chemical Physics of Solids, N{\"o}thnitzer Str. 40, 01187 Dresden, Germany}
\author{Sahana R\"{o}ßler}
    \affiliation{Max Planck Institute for Chemical Physics of Solids, N{\"o}thnitzer Str. 40, 01187 Dresden, Germany}
\author{Manuel Valvidares}
    \affiliation{ALBA Synchrotron Light Source, Cerdanyola del Valles, Barcelona 08290, Spain}
\author{Maurits~W.~Haverkort}
     \affiliation{Institute for Theoretical Physics, Heidelberg University, Philosophenweg 19, 69120 Heidelberg, Germany}
\author{Yu Liu}
    \altaffiliation{Center for Correlated Matter and Department of Physics, Zhejiang University, Hangzhou 310058, China}
    \affiliation{Los Alamos National Laboratory, Los Alamos, New Mexico 87545, USA}
\author{Eric D. Bauer}
    \affiliation{Los Alamos National Laboratory, Los Alamos, New Mexico 87545, USA}
\author{Liu~Hao~Tjeng}
    \affiliation{Max Planck Institute for Chemical Physics of Solids, N{\"o}thnitzer Str. 40, 01187 Dresden, Germany}
\author{Gertrud~Zwicknagl}
    \affiliation{Technische Universit{\"a}t Braunschweig, Braunschweig, Germany}
    \affiliation{Max Planck Institute for Chemical Physics of Solids, N{\"o}thnitzer Str. 40, 01187 Dresden, Germany}
\author{Andrea~Severing}
        \affiliation{Institute of Physics II, University of Cologne, Z\"{u}lpicher Str. 77, 50937 Cologne, Germany}
        \affiliation{Max Planck Institute for Chemical Physics of Solids, N{\"o}thnitzer Str. 40, 01187 Dresden, Germany}

\date{\today}

\begin{abstract}
CeRh$_2$As$_2$ is a new multiphase superconductor with strong suggestions for an additional itinerant multipolar ordered phase. The modeling of the low temperature properties of this heavy fermion compound requires a quartet Ce$^{3+}$ crystal-field ground state. Here we provide the evidence for the formation of such a quartet state using x-ray spectroscopy.  Core-level photoelectron and x-ray absorption spectroscopy confirm the presence of Kondo hybridization in CeRh$_2$As$_2$. The temperature dependence of the linear dichroism unambiguously reveils the impact of Kondo physics for coupling the Kramer's doublets into an effective quasi-quartet.  Non-resonant inelastic x-ray scattering data find that the $|\Gamma_7^- \rangle$ state with its lobes along the 110 direction of the tetragonal structure ($xy$ orientation) contributes most to the multi-orbital ground state of CeRh$_2$As$_2$.

\end{abstract}


\maketitle

Characterizing novel ordering and understanding how they arise is key for the development of functional materials. Heavy-fermion materials have been important model systems for the discovery of novel phases as well as their cooperation, coexistence, or competition. This is due to the fact that the characteristic energy scale of these materials is of the order of a few $meV$ which, in turn, implies a high tunability of the electronic properties\,\cite{Floquet2005,Thalmeier2005,Fulde2006,Coleman2007,Khomskii2010}. CeRh$_{2}$As$_{2}$ is a heavy-fermion system of high current interest, experimentally\,\cite{Khim2021, kimura2021,Kibune2022,Hafner2022,Landaeta2022a,Mishra2022} as well as theoretically\,\cite{Mockli2021a, Schertenleib2021, Ptok2021, Skurativska2021, Nogaki2021, Mockli2021b, Cavanagh2022, Nogaki2022,  Machida2022, Hazra2023}. This material exhibits multiphase unconventional superconductivity (SC)\,\cite{Pourret2021,Nica2022} and putative itinerant multipolar order; there are indications for a quadrupolar density wave (QDW) below $T_0$\,=\,0.48\,K and above the superconducting transition at $T_{SC}$\,$\approx$\,0.31\,K; the transition temperatures refer to newer sample generations\,\cite{Semeniuk2023}. The unusual behavior of CeRh$_{2}$As$_{2}$ is due to the presence of $4f$ electrons. This conclusion is derived from the properties of its non-$f$ reference compound LaRh$_{2}$As$_{2}$ that can be quantitatively described by standard Eliashberg theory of strong-coupling electron-phonon superconductors\,\cite{Landaeta2022b}. For understanding the low temperature ($T$) phases of CeRh$_2$As$_2$ it is also crucial that its $f$ electrons form a quasi-quartet ground state.

In the heavy fermion state, the 4$f$ degrees of freedom form heavy quasiparticles due to the Kondo effect ($cf$-hybridization). CeRh$_2$As$_2$ forms in the globally centro-symmetric tetragonal CaBe$_2$Ge$_2$ structure with two identical Ce sites per unit cell. The Ce sites lack inversion symmetry so that the local point symmetry is C$_{4v}$ but the crystal-field split Ce$^{3+}$ Hund's rule ground state of Ce$^{3+}$ may be treated like D$_{4h}$ because the contribution of 5$d$ states is negligible. Accordingly, the three Kramer's doublets can be written in $J_z$ formalism as $|\Gamma_7^- \rangle$\,=\,$\sqrt{1-\alpha^2} \cdot |\pm 3/2 \rangle$\,-\,$|\alpha| \cdot |\mp 5/2 \rangle$, $|\Gamma_6\rangle$\,=\,$|\pm1/2 \rangle$, $|\Gamma_7^+\rangle$\,=\,$\sqrt{1-\alpha^2} \cdot |\pm 5/2 \rangle$\,+\,$|\alpha| \cdot |\mp 3/2 \rangle$, and the $f$ contribution to the quasi-particles has the character of the crystal-field ground state when the excited states are well separated from the ground state. However, if the system presents a Kondo temperature $T_K$ of the order of magnitude of crystal-field splittings energies $\Delta_i$, one may expect the Kondo effect to induce an effective quasi-quartet ground state. Its character reflects both states of the quasi-quartet.

The conjecture of a Kondo induced quasi-quartet is at the heart of renormalized band structure calculations in Ref.\,\cite{Hafner2022} that find superconductivity and other types of order may exist at different sites of the Fermi surface. Here, we investigate the electronic structure of Ce in CeRh$_2$As$_2$ in order to find spectroscopic evidence for such a quasi-quartet. Core-level photoelectron spectroscopy with hard x-rays (HAXPES) is sensitive to covalence and hence to the filling $n_f$ of the Ce\,4$f$ shell\,\cite{Gunnarsson1983,Fuggle1983,Gunnarson2001,Sundermann2016}. Also x-ray absorption spectroscopy (XAS) at the Ce\,$M_{4,5}$ edges can reveal signatures of hybridization effects by exhibiting, in addition to the main transition $3d^{10}4f^1 \rightarrow 3d^94f^2$, a satellite induced by hybridization with the $3d^{10}4f^0$ configuration\,\cite{Gunnarsson1983,Fuggle1983,Willers2012a}. The linear dichroism (LD) of XAS at the Ce\,$M_{4,5}$ edges, on the other hand, is sensitive to the symmetry of the crystal-field ground state when exploiting the dipole selection rules of linear polarized light\,\cite{Hansmann2008, Willers2012a}; at low $T$ the ground state is probed and excited states contribute to the net dichroism when they are populated thermally. This way the $T$-dependence of the LD provides insight into the sequence of crystal-field states, and size of the crystal-field splitting energies $\Delta_{1,2}$. Non-resonant inelastic x-ray scattering (NIXS) is also sensitive to the crystal-field ground state but it is based on multipole selection rules\,\cite{Schuelke2007,Gordon2008,Gordon2009,Willers2012b,Sundermann2019}. In NIXS, the direction of the momentum transfer with respect to the sample orientation takes over the role of the electric field vector in XAS. In contrast to XAS, NIXS is also sensitive to anisotropies with fourfold rotational symmetry, so that the orientation of e.g. a $|\Gamma_7^{+/-}\rangle$ charge density within the tetragonal unit cell can be determined.

In the present letter, we report results of HAXPES as well as XAS and NIXS measurements of CeRh$_2$As$_2$ single crystals. The spectra are analyzed with a full multiplet calculation to gain insight into the electronic structure of the crystal-field split Hund's rule ground state of Ce$^{3+}$. We use a simplified Anderson impurity calculation (SIAM) in the non-crossing approximation (NCA) to show that in the presence of the Kondo effect and in case $T_K$\,$\approx$\,$\Delta_i$, the excited crystal-field states do contribute to the LD also at low temperatures. 

\begin{figure}[]
    \includegraphics[width=0.85\columnwidth]{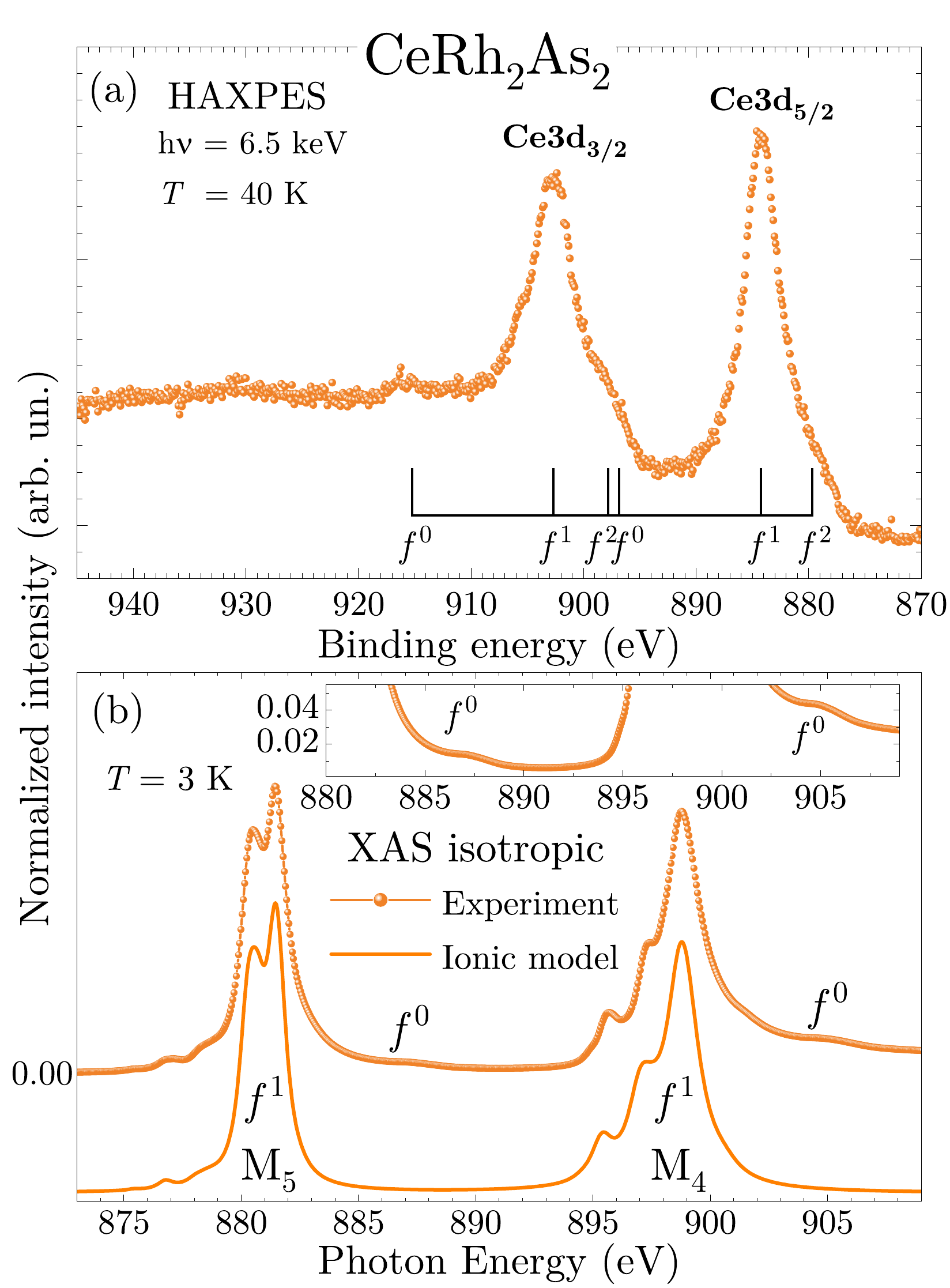}
\vspace{-0.3cm}
    \caption{(a) Ce 3$d$ core level HAXPES data of CeRh$_2$As$_2$. The black ruler at the bottom indicates the position of the $f^0$, $f^1$ and $f^2$ spectral weights. (b) Isotropic XAS spectrum of CeRh$_2$As$_2$ (dots) and the corresponding ionic multiplet simulation of the $f^1$ configuration (solid line). The inset shows a close-up to the energy regions with the $f^0$ satellites.}
    \label{fig1}
\end{figure}

Growth and characterization of single crystalline CeRh$_2$As$_2$ are described in the Appendix. The transition temperatures are $T_{SC}$\,=\,0.30\,K and $T_0$\,=\,0.45\,K, which is in very good agreement with the values reported for newer generations of samples\,\cite{Semeniuk2023}.

XAS spectra at the Ce $M_{4,5}$ edges (880-904\,eV) were measured in the total electron yield mode (TEY) at beamline BL29 BOREAS at ALBA synchrotron, Spain\,\cite{Barla2016} with an energy resolution of 300\,meV. Clean sample surfaces were obtained by cleaving the CeRh$_2$As$_2$ single crystals  \textit{in-situ} in an ultra-high-vacuum (UHV) of about 1x10$^{-9}$\,mbar prior to inserting them into the main chamber with a base pressure of 10$^{-10}$\,mbar. XAS data were taken at 3, 50, 100, and 200\,K with the electric field $\vec{E}$\,$\parallel$\,$ab$ and $\vec{E}$\,$\parallel$\,$c$.  NIXS experiments were performed at the Max-Planck NIXS end station P01 at PETRA III/DESY, Germany at 8\,K with the same set-up as described in Ref.\,\cite{Sundermann2017}. The average momentum transfer amounted to $|\vec{q}|$\,=\,9.6\,$\pm$\,0.1\,\AA$^{-1}$. The $N_{4,5}$ edges were measured with the momentum transfer $\vec{q}$ parallel to [100], to [110], and [001]. HAXPES experiments were carried out at the Max-Planck-NSRRC HAXPES end station at the Taiwan undulator beam line BL12XU at SPring-8, Japan with a photon energy of about 6.5 keV, an overall energy resolution of 250\,meV, and a sample temperature of 40\,K\,\cite{Weinen2015}. Samples were polished \textit{in-situ} in order to expose a fresh layer, and wide-scans were performed to ensure the absence of O and C 1$s$ signal from surface contamination or oxidation. The pressure in the main chamber was in the low 10$^{-10}$\,mbar. More information about the experimental set-ups can be found in the Appendix.

XAS and NIXS data are simulated with the full multiplet code $Quanty$\,\cite{Haverkort2016}, starting with atomic parameters from the Cowan code\,\cite{CowanBook}, and applying typical reductions for the respective Hartree-Fock values of the Slater integrals and spin orbit values\,\cite{Tanaka1994} (see Appendix).  The reduction factors were optimized to best reproduce the isotropic spectrum in the case of XAS\,\cite{Hansmann2008} and (pseudo)isotropic for NIXS. Both are constructed from the directional dependent data ($I_{iso}$\,=\,(2\,I$_{\parallel ab}$\,+\,I$_{\parallel c}$)\,/3). Furthermore, a Gaussian and Lorentzian broadening are applied to account for resolution and lifetime effects (see Appendix).

Figure\,\ref{fig1}(a) shows the Ce\,$3d$ core level HAXPES data of CeRh$_2$As$_2$ without any background correction. CeRh$_2$As$_2$ shows the three spectral weights that arise in the presence of the core hole, when the ground state is a coherent interference of the 4$f^0$, 4$f^1$, and 4$f^2$ configurations\,\cite{Gunnarson2001}. Also the \textit{isotropic} XAS spectrum, shown in Fig\,\ref{fig1}\,(b), reveals the presence of a 4$f$ configuration different from 4$f^1$. The spectral weights corresponding to the 3$d^{10}$4$f^0$\,$\rightarrow$\,3$d^{9}$4$f^1$ transition show up as small satellites on the high energy tails of the $M_5$ and $M_4$ absorption lines, shown on an enlarged scale in the inset.  These small intensities originating from the 4$f^0$ configuration confirm that the Kondo effect is indeed present in CeRh$_2$As$_2$, an aspect that turns out to be crucial for the analysis of the temperature dependence of the linear dichroism in the XAS spectra as discussed below. 

\begin{figure}[]
\centering
    \includegraphics[width=0.85\columnwidth]{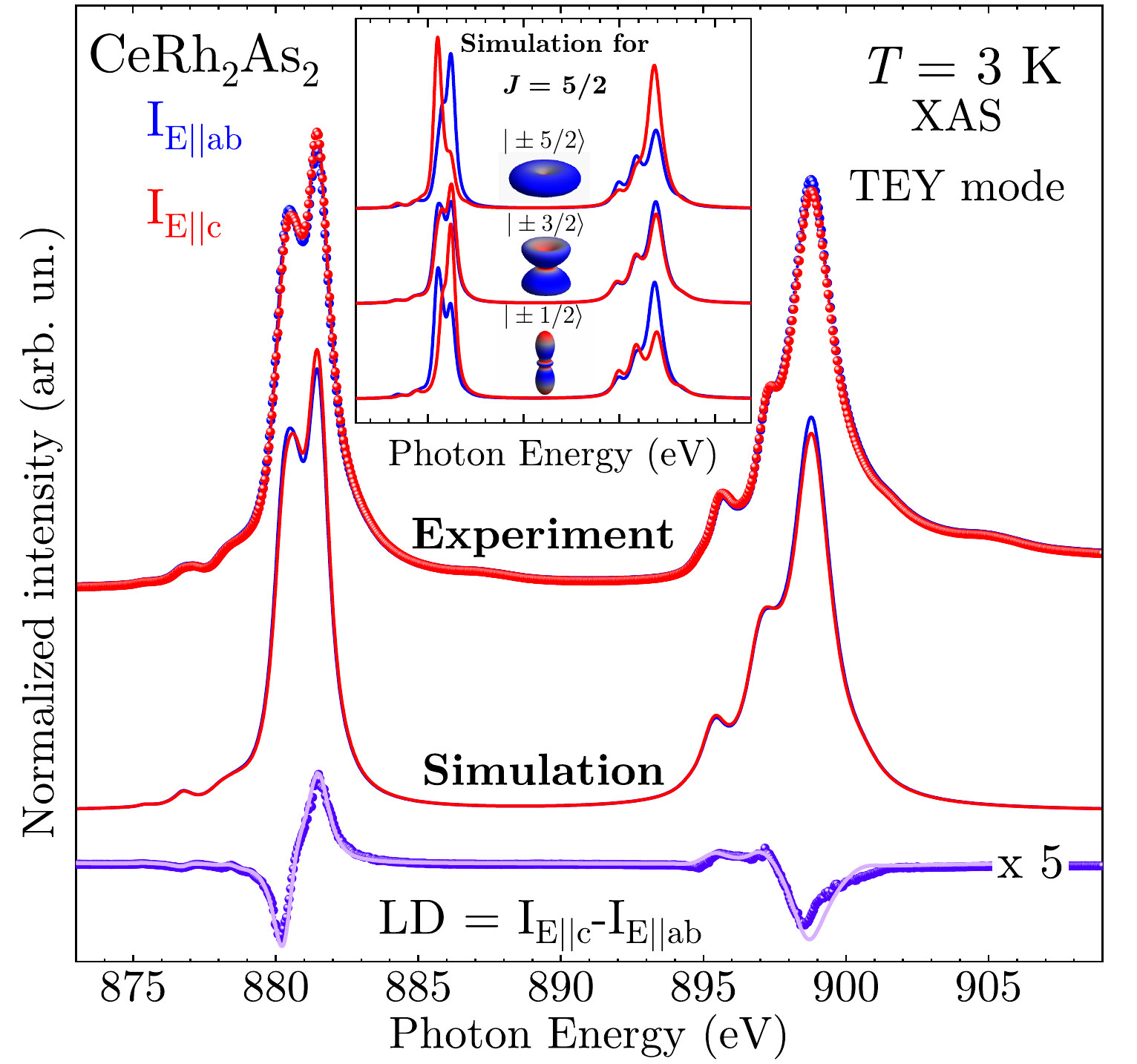}
\vspace{-0.3cm}
    \caption{Linear polarized XAS spectra of CeRh$_2$As$_2$ at 3\,K for the electric field vector $\vec{E}$\,$\parallel$\,$c$ (red) and $\vec{E}$\,$\parallel$\,$ab$ plane (blue), and linear dichroism (purple), dots for the experiment and lines are the full multiplet simulation of the $f^1$ configuration. The inset shows the calculated individual contributions to each pure $|\pm J_z \rangle$ state. }
    \label{fig2}
    \vspace{-0.2cm}
\end{figure}

\begin{figure}[t]
	\includegraphics[width=0.85\columnwidth]{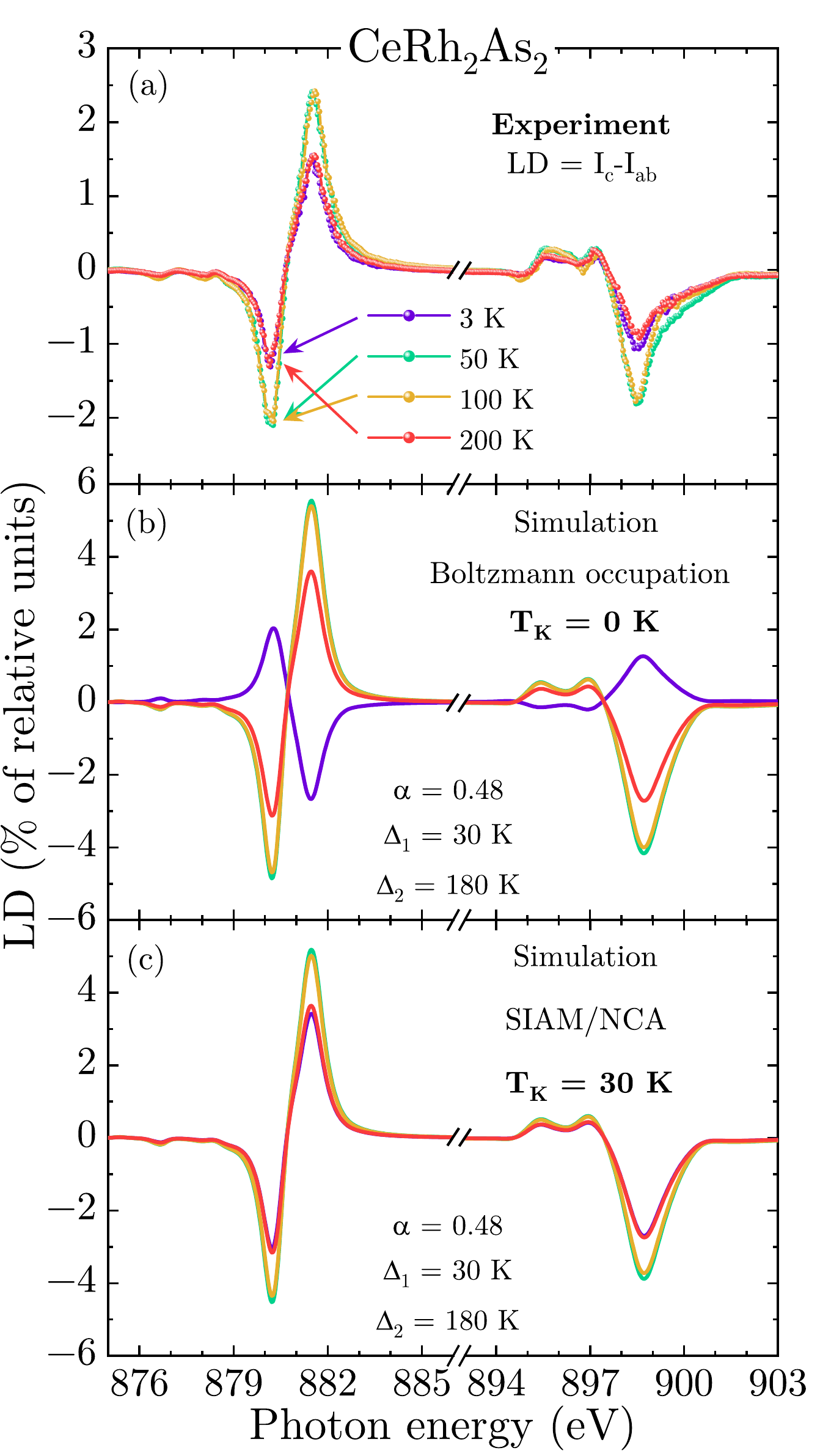}
	\vspace{-0.3cm}
	\caption{(a) Measured $T$-dependence of the LD at $T$\,=3 (purple), 50 (green), 100 (yellow), and 200\,K (red). The panels below show two different simulations, an ionic crystal-field calculation considering the Boltzmann occupation of excited states but no Kondo effect (b), and the same crystal-field model including the Kondo effect (c), see text. The scale was multiplied by 100 to show the percentage of LD.}
	\label{fig3}
	\vspace{-0.5cm}
\end{figure}

The top spectra of Fig.\ref{fig2} show the polarization dependent XAS data of CeRh$_2$As$_2$ at 3\,K  (red and blue circles) and at the bottom the experimental linear dichroism LD\,=\,I$_{E||c}$\,-\,I$_{E||ab}$ 5 times enlarged (purple circles). At first, we compare the data with the simulations of the pure $|\pm J_z \rangle$ states of the $f^1$ configuration that are displayed in the inset. Obviously a pure $|\pm 5/2 \rangle$  and $|\pm 1/2 \rangle$ can be excluded as ground state. The pure $|\pm 3/2 \rangle$ may resemble the data best, but as we will explain below, we will need not only to consider the correct crystal-field scheme but also the Kondo effect. We note that the measured dichroic signal is relatively small and we verify its reliability by analyzing the lineshape. We recall that the lineshape of the $M_{4,5}$ dichroism of Ce$^{3+}$ ions is unique as long as the crystal-field splitting is negligible compared to the inverse lifetime of the $M_{4,5}$ white lines, see Ref.s\,\cite{Sundermann2015,Amorese2020}. After optimizing the reduction factors by means of the isotropic spectrum (see Fig.\ref{fig1}(b)), we plot this unique lineshape on top of the measured dichroic spectrum and we indeed conclude that the experimental features all belong to a Ce$^{3+}$ ion in the small crystal-field limit. Also the corresponding simulations for the $M_{4,5}$ spectra match the measured ones excellently . 

\begin{figure*}[t]
	\begin{center}
		\includegraphics[width=1.8\columnwidth]{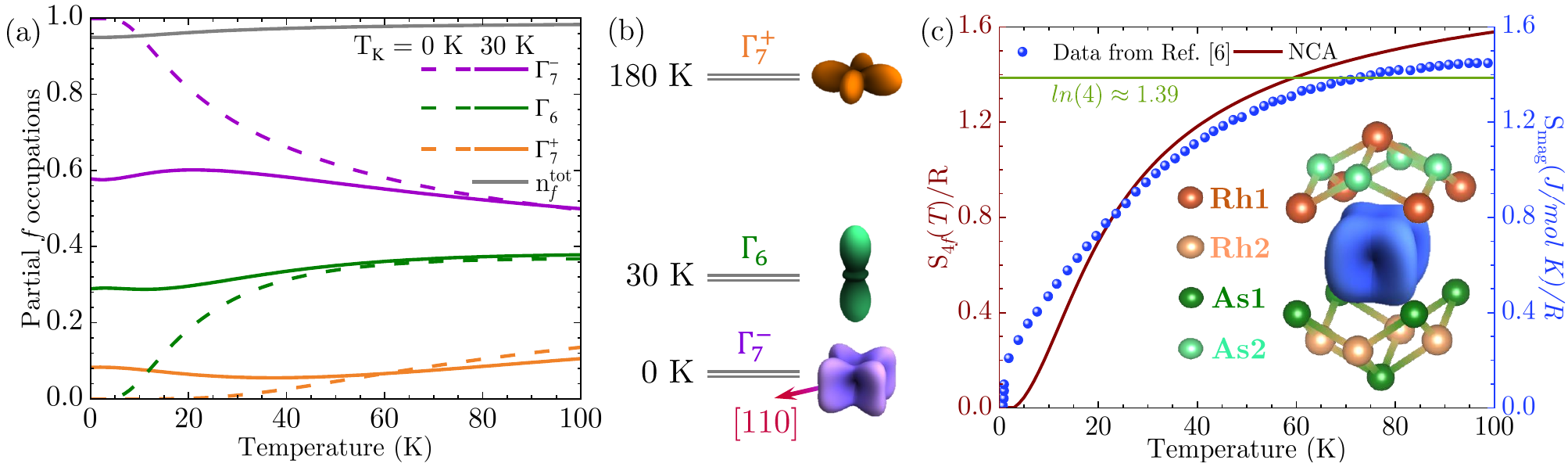}
		\vspace{-0.8cm}
	\end{center}
	\caption{(a) SIAM/NCA calculation of partial 4$f$ occupations for $T_K$\,=\,30\,K (solid lines) and Boltzmann statistics ($T_K$\,=\,0 - dashed lines) for the crystal-field scheme according Ref.\cite{Hafner2022}, shown in (b), with $|\Gamma_7^- \rangle$\,=\,0.88\,$|\pm 3/2 \rangle$\,-\,0.48\,$|\mp 5/2 \rangle$, $|\Gamma_6\rangle$\,=\,$|\pm1/2 \rangle$, $|\Gamma_7^+\rangle$\,=\,0.88\,$|\pm 5/2 \rangle$\,+\,0.48\,$|\mp 3/2 \rangle$. (c) SIAM/NCA calculated entropy as function of temperature (red solid line) compared to specific heat data from Ref.\,\cite{Khim2021} (blue dots) and expected entropy of a quartet (horizontal green line). Inset: Ce charge density of the Kondo-mixed ground state (58\%$|\Gamma_7^- \rangle$, 29\% $|\Gamma_6 \rangle$, and 8\% $|\Gamma_7^+ \rangle$) with its nearest neighbors.}
	\label{fig4}
\end{figure*}

The temperature dependence of the LD($T$) at the $M_5$ and $M_4$ edges is shown on an expanded energy scale in Fig.\,\ref{fig3}(a) for 3, 50, 100, and 200\,K. The LD at 50\,K has increased with respect to 3\,K, as indicated by the arrows. The change in LD($T$) indicates that an excited crystal-field state is close. LD($T$) remains almost as large at 100\,K, and then decreases when warming up to 200\,K. Also here it is informative to understand the $T$-dependence of the LD by looking at first at the pure $|\pm J_z \rangle$ states that are depicted in the inset of Fig.\ref{fig2}. The $|\pm 1/2 \rangle$ has a stronger dichroism of the same sign as the one at 3\,K whereas the LD of the $|\pm 5/2 \rangle$ state is large and opposite. Hence, the increase of the LD($T$) when warming up to 50\,K must be due to the increasing population of the $|\Gamma_6\rangle$\,=\,$|\pm 1/2 \rangle$ state, followed by the successive population of the second $|\Gamma_7^{-/+}\rangle$ state of majority $|\pm 5/2 \rangle$ that compensates the LD of the $|\Gamma_6\rangle$ when warming up further to 100 and 200\,K.

First, we simulate LD($T$) by assuming the crystal-field model that satisfies the macroscopic data best, especially the high temperature anisotropy of the static susceptibility where the impact of the Kondo effect can be neglected. Hafner \textit{et al}.\cite{Hafner2022} propose the sequence of states that we obtain from the above comparison with the pure $|J_z\rangle$ states, with $\Delta_1$\,=30\,K, $\Delta_2$\,=180\,K and $\alpha$\,=\,0.48 for the crystal-field splittings and mixing factor of the ground state, respectively. These numbers are based on the analysis of the specific heat and static susceptibility. This model reproduces qualitatively LD($T$) at 50, 100, and 200\,K when taking into account the respective Boltzmann occupations of the excited states (see Fig.\,\ref{fig3}(b)). It contradicts, however, the experiment at low temperatures where the Kondo effect starts to manifest: at 3\,K the simulated LD turns out to be opposite to the experimental one (compare Fig.\,\ref{fig3}(a) and (b)).

We take the discrepancy between data and crystal-field simulation of LD($T$) at low $T$ as a strong suggestion for having to consider the impact of the Kondo interaction since Kondo is a low $T$ effect. The SIAM/NCA calculations\,\cite{Zevin1988,Zwicknagl1990,Amorese2020} take into account the hybridization of the three crystal-field states with the conduction electron bath. It turns out that the Kondo effect induces a finite occupation of the first and second excited crystal-field states also at 3\,K, in contrast to the Boltzmann occupations where only the lowest state is occupied. The Kondo induced (solid lines) and Boltzmann-only (dashed lines) occupations as function of $T$ are displayed in Fig.\,\ref{fig4}\,(a). The SIAM/NCA calculations take into account the same crystal-field model as above and a Kondo temperature $T_K$ of 30\,K. The latter is suggested from macroscopic data\,\cite{Khim2021, Hafner2022}. The Kondo-mixed ground state at low $T$ is calculated with a 4$f$-shell occupation of $n_f^{tot}$\,=\,95\%. Beginning at 50\,K, but certainly at 100\,K the differences between Kondo induced and Boltzman-only occupation are negligible. Therefore, the Kondo effect has the strongest impact on the simulated LD($T$) at 3\,K: due to the Kondo induced contribution of the higher lying states, the sign of the LD at 3\,K is now correctly reproduced (see Fig.\,\ref{fig3}(c)). 

The calculation of the entropy within the SIAM/NCA model based on the above crystal-field scheme and Kondo temperature yields the red line in Fig.\,\ref{fig4}\,(c), which reproduces $R$\,ln4 at about 60\,K very well, and it is in good agreement with the analysis of the experimental specific heat (blue dots) by Khim \textit{et al.}\,\,\cite{Khim2021}. 

While a $T_K$ of 30\,K reproduces LD($T$) well, we show in the Appendix analyses with other $T_K$ values. We find that a lower value of $T_K$\,=\,15\,K still produces the wrong sign in the LD at 3\,K, see Fig.\,\ref{fig6}]. This indicates that $k_B$$T_K$ must be at least of the same order of magnitude as the crystal-field splitting, thus implying that the degeneracy of the relevant low-energy subspace is effectively higher than two-fold. We also find that the entropy sets limits on the upper and lower value of $T_K$, see Fig.\,\ref{fig7}.

\begin{figure}[t]
    \includegraphics[width=0.80\columnwidth]{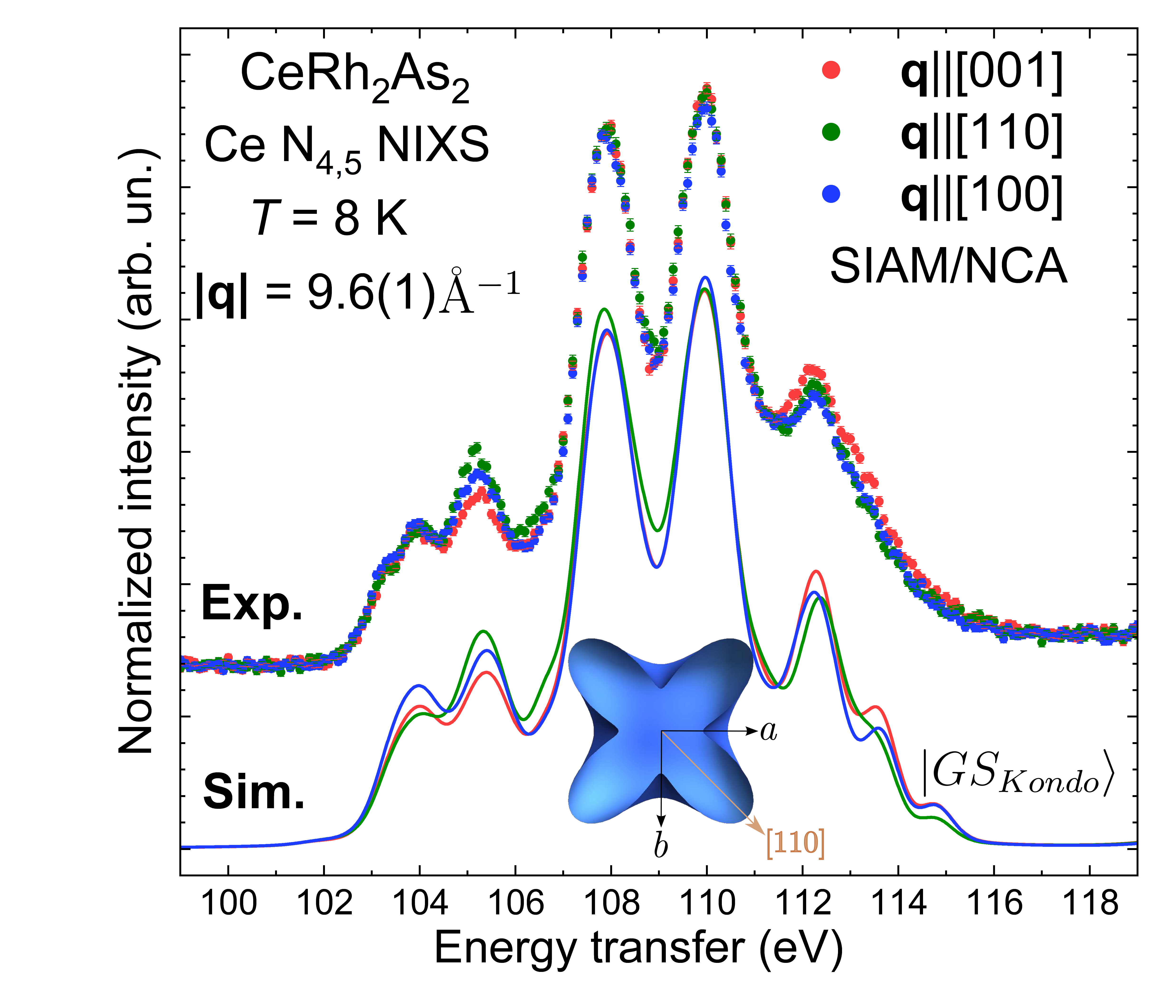}
\vspace{-0.3cm}
    \caption{Normalized and background corrected NIXS data for $\mathbf{q}\,\parallel[001]$\,(red), $\mathbf{q} \parallel[110]$\,(green) and $\mathbf{q} \parallel[100]$\,(blue). The lines represent the simulation in the SIAM/NCA work-frame for $|GS_{Kondo}\rangle$ ground state, with the $|\Gamma_7^-\rangle$ most strongly occupied. The inset represent the respective orbital orientation within the $ab$-plane.}
    \label{fig5}
    \vspace{-0.5cm}
\end{figure}

LD($T$) is well described with a $|\Gamma_7^{+/-}\rangle$ lowest in energy. It is, however, not yet clear whether it is the $|\Gamma_7^{+}\rangle$ ($x^2$\,-\,$y^2$) or the $|\Gamma_7^{-}\rangle$ ($xy$ orientation). We address this question with directional dependent NIXS. NIXS spectra at 8\,K for momentum transfers $\vec{q}$\,$\parallel$\,(100), $\vec{q}$\,$\parallel$\,(110), and for completeness also for $\vec{q}$\,$\parallel$\,(001), are displayed in Fig.\,\ref{fig5}. The spectra are normalized to the integral of the Compton background at higher energy transfers, and a linear background is subtracted as described in the Appendix and displayed in Fig.\,\ref{fig8}. The directional dependence in the $ab$-plane is fairly weak, except for a more distinct region at around 105\,eV energy transfer. The in-plane directional dependence  gives the desired information about the sign of the $|\Gamma_7^{+/-}\rangle$ state. The 45$^{\circ}$ rotation would imply a switch in intensity of the (100)(blue) and (110)(green) signals. Also the NIXS spectra were simulated with the Kondo mixed ground state $|GS_{Kondo}\rangle$, using the respective occupations at 8\,K. The Kondo mixing of states reduces and alters slightly the directional dependence in the $ab$-plane with respect to pure $|\Gamma_7^{+/-}\rangle$ states (see Fig.\,\ref{fig8} in the Appendix) due to the admixture of the rotational invariant $|\Gamma_6\rangle$ and the excited $|\Gamma_7^{-/+}\rangle$ state with lobes rotated by 45$^{\circ}$ and, consequently, an opposite in-plane anisotropy. The resulting NIXS simulation with the orbital mixed ground state $|GS_{Kondo}\rangle$ shows that the lobes are along (110), thus confirming the $|\Gamma_7^{-}\rangle$ state must be the one most strongly occupied. A more detailed discussion about the NIXS spectra is found in the Appendix.

Putting together the results of macroscopic data, and the present $T$ dependent XAS and low $T$ NIXS, we find that the 4$f^1$ part of the multi-orbital ground state wave function with $n_f^{tot}$=0.95 consists of $\sqrt{0.58}\,|\Gamma_7^-\rangle$, $\sqrt{0.29}\,|\Gamma_6\rangle$, and $\sqrt{0.08}\,|\Gamma_7^+\rangle$ with $|\Gamma_7^-\rangle$ = 0.88$|\pm 3/2 \rangle$\,-0.\,48$|\mp 5/2 \rangle$, $|\Gamma_7^+\rangle$ = 0.48$|\pm 3/2 \rangle$\,+0.\,88$|\mp 5/2 \rangle$, and $\Gamma_6\rangle$\,=\,$|\pm 1/2 \rangle$. The corresponding charge density is displayed in Fig.\,\ref{fig4}\,(b) and Fig.\,\ref{fig5}.  The dominance of the $|\Gamma_7^-\rangle$ state accommodates very well the Fermi surface as calculated in Ref.\,\cite{Hafner2022}.

Quantitatively, the present simulation overestimates the $T$-dependence of LD($T$). This is most likely due to the fact that the present SIAM/NCA model is too simplistic. For example, the calculations performed on the NCA frame are in the limit U\,$\rightarrow$\,$\infty$, meaning they do not take into account the influence of the 4$f^2$ configuration and all weight is concentrated on 4$f^1$. Moreover, symmetry dependent hybridization is not taken into account, although renormalized band structure calculations that do consider symmetries\,\cite{Zwicknagl1992} show that the $|\Gamma_7^+\rangle$ at 180\,K contributes significantly to the Fermi surface; actually more than expected from an isotropic calculation (see Fig.\,\ref{fig10} in the Appendix).

To summarize, the presence of significant $cf$-hybridization in CeRh$_2$As$_2$ is consistent with core level HAXPES and $M$-edge XAS data. The temperature dependence of the linear dichroism LD($T$) in $M$-edge XAS is qualitatively reproduced with the crystal-field model that was suggested from the anisotropy of the static susceptibility and the specific heat when taking into account the Kondo induced coupling of all crystal-field states to the conduction electron bath. The resulting multi-orbital ground state is an effective quartet that reproduces well the asymptotic value of entropy from the specific heat, thus validating the conjecture of a quasi-quartet ground state that is crucial for superconductivity to coexist with other types of order.

\section{acknowledgement} All authors thank Manuel Brando, Elena Hassinger, Javier Landaeta, and Konstantin Semeniuk for enlightening discussions. Katharina H{\"o}fer's expertise in designing the cleaving setup was extremely helpful and is appreciated by all authors. A.S and M. F.-C. benefited from support of the German Research Foundation (DFG), Project No. 387555779. Y. L. and E.D.B. were supported by the U.S. Department of Energy (DOE), Office of Basic Energy Sciences, Division of Materials Science and Engineering under project "Quantum Fluctuations in Narrow-Band Systems". XAS and XLD measurements were performed at ALBA under proposal ID 2022097004. We acknowledge DESY (Hamburg, Germany), a member of the Helmholtz Association HGF, for the provision of experimental facilities. 


\section{Appendix}

	\subsection{Sample growth and characterization}
		Single crystals of CeRh$_2$As$_2$ were grown in Bi flux starting from a mixture of pure elements Ce, Rh, As, and Bi with a molar ratio of 1 : 2 : 2 : 30. The starting materials were sealed in an evacuated fused silica tube, which was heated to 1150$^{\circ}$C over 30\,h, followed by a dwell at 1150$^{\circ}$C for 24\,h. A \textit{sawtooth} heating/cooling profile was used to produce larger and higher-purity crystals consisting of:  cooling from 1150$^{\circ}$C to 750$^{\circ}$C at a rate of 20$^{\circ}$C/hr., heating to 1050$^{\circ}$C  at 75$^{\circ}$C/hr., cooling to 750$^{\circ}$C at 7.5$^{\circ}$C/hr., heating to 1050$^{\circ}$C at 75$^{\circ}$C/hr., then slowly cooling to 700$^{\circ}$C at 1.9$^{\circ}$C/hr.
	
	The T$_{SC}$ and $T_0$ of 0.30 and 0.45\,K are in excellent agreement with the higher quality CeRh$_2$As$_2$ crystals from Semeniuk \textit{et al.}\,\cite{Semeniuk2023}. The peak height and sharpness of the transition at $T_{SC}$ and $T_0$ in the specific heat, and a residual resistivity of RRR\,=\,2.4, are all indications of improved crystal quality\,—\,all comparable to the Semeniuk samples.
	
	The morphology of the CeRh$_2$As$_2$ single crystals was very well defined, presenting as thin platelets with the [001] direction normal to the plane, and their in-plane axes typically aligned to natural edges of the crystals.
	
	\subsection{Experimental details}
	The single-crystalline samples were aligned using the Laue method, and mounted according to the requirements of each technique. For cleaving the small crystals in the XAS experiments, minute grooves were machined into the sample holders where the crystals were glued, to provide more support to the stress of cleaving. Similarly, small custom-made posts were developed with a recess to fit into each sample and assure better grip. These precautions were needed because it turned out that the crystals were very hard to cleave. The single-crystalline samples were aligned so that a surface with a [100] normal vector was exposed to the beam. This way, both polarizations, $\vec{E}$\,$\parallel$\,$ab$ and $\vec{E}$\,$\parallel$\,$c$, could be measured without rotating the sample. For NIXS the samples were mounted making use of the flat and mirror-like surfaces after aligned properly, so that the momentum transfers $\vec{q}$ parallel to [100], to [110], and [001] could be realized. In the HAXPES experiment the photons are horizontally polarized. The MB Scientific A-1 HE analyzer was mounted horizontally at 90$^{\circ}$\,\cite{Weinen2015}, and the sample was measured with the [001] direction 15$^{\circ}$\,away from the analyzer (near normal emission).
	
	\subsection{Simulations}
	The reduction of the atomic parameters accounts for configuration interaction effects that are not included in the Hartree-Fock scheme\,\cite{Tanaka1994}. For the XAS simulations, the 4$f$-4$f$  and 3$d$-4$f$ Slater integrals, and the atomic value for the 3$d$ spin-orbit were reduced by 39.0(5)\% 17.5(5)\%, and 2.7(5)\%, respectively\,\cite{Hansmann2008}. The reduction factors were optimized to best reproduce the isotropic spectrum that was constructed from the directional dependent data ($I_{iso}$\,=\,(2\,I$_{\parallel ab}$\,+\,I$_{\parallel c}$)\,/3). A Gaussian broadening for resolution and an energy dependent Lorentzian broadening for life time effects was used.
	
	For the NIXS simulations the resolution Gaussian shaped broadening of 0.7\,eV and an additional Lorentzian broadening of 0.17\,eV FWHM for life-time effects were taken into account. Here the Slater integrals for the 4$f$-4$f$ and 4$d$-4$f$ Coulomb interactions were reduced by about 26(1)\,\% and 22(1)\,\%, respectively. The spin-orbit reduction associated to the 4$d$ was reduced by 2(1)\,\%, and in particular, the Coulomb exchange term $G^1$, that controls the dipolar contribution's energy transfer, was reduced by 35.5(5)\,\%. Also here the reduction factors were optimized by reproducing the energy distribution of the (pseudo)isotropic Ce $N_{4,5}$ data that were, as for XAS, constructed from the directional dependent spectra.
	
	\begin{figure}[]
		\includegraphics[width=0.9\columnwidth]{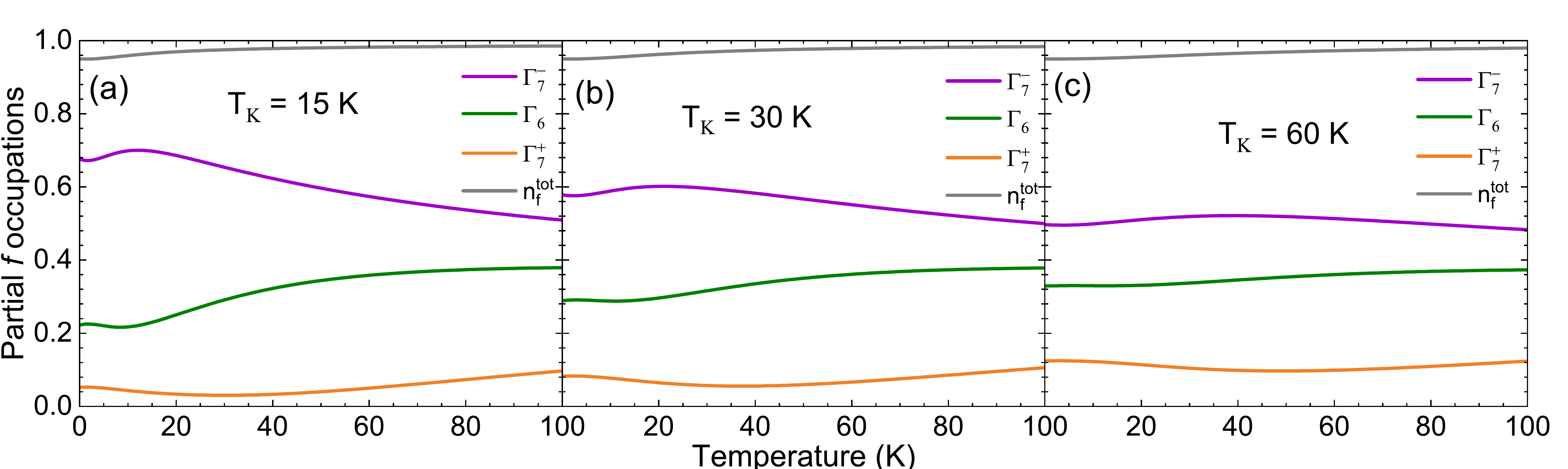}
		\includegraphics[width=0.9\columnwidth]{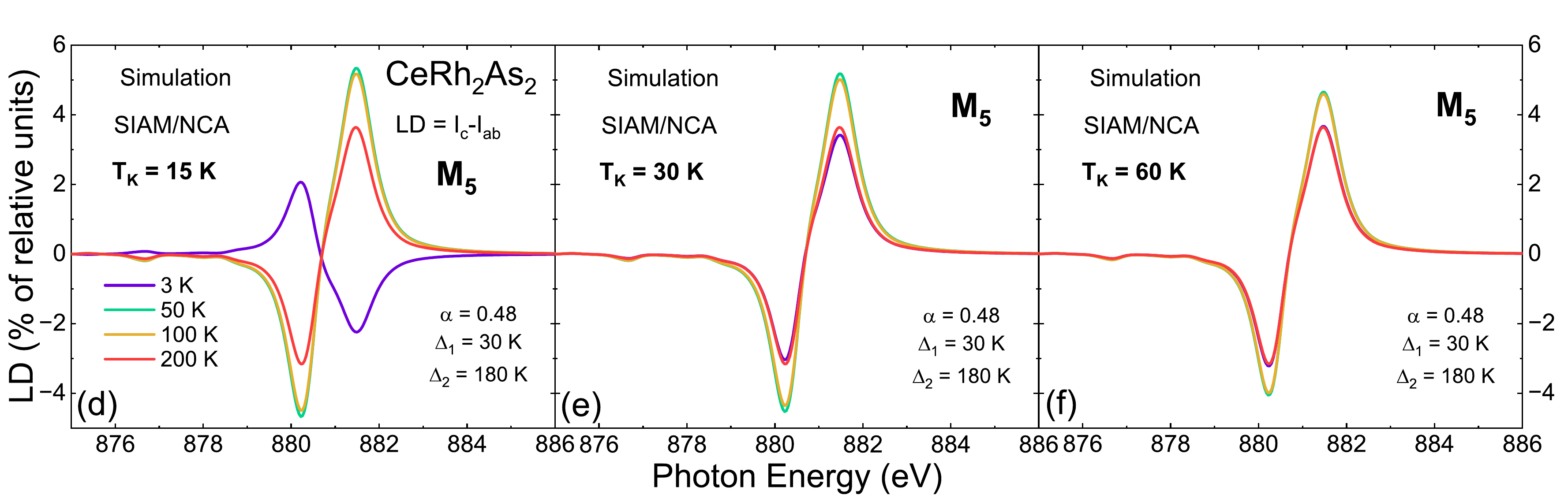}
		\caption{(a)-(c) Partial occupation of $f$ states according to SIAM/NCA calculation and (d)-(f) resulting LD($T$) for $T_K$\,=15, 30, and 60\,K.}
		\label{fig6}
	\end{figure}
	
	\subsection{Orbital occupation, LD($T$) and entropy for different values of $T_K$}
	
	\begin{figure}[]
		\includegraphics[width=0.6\columnwidth]{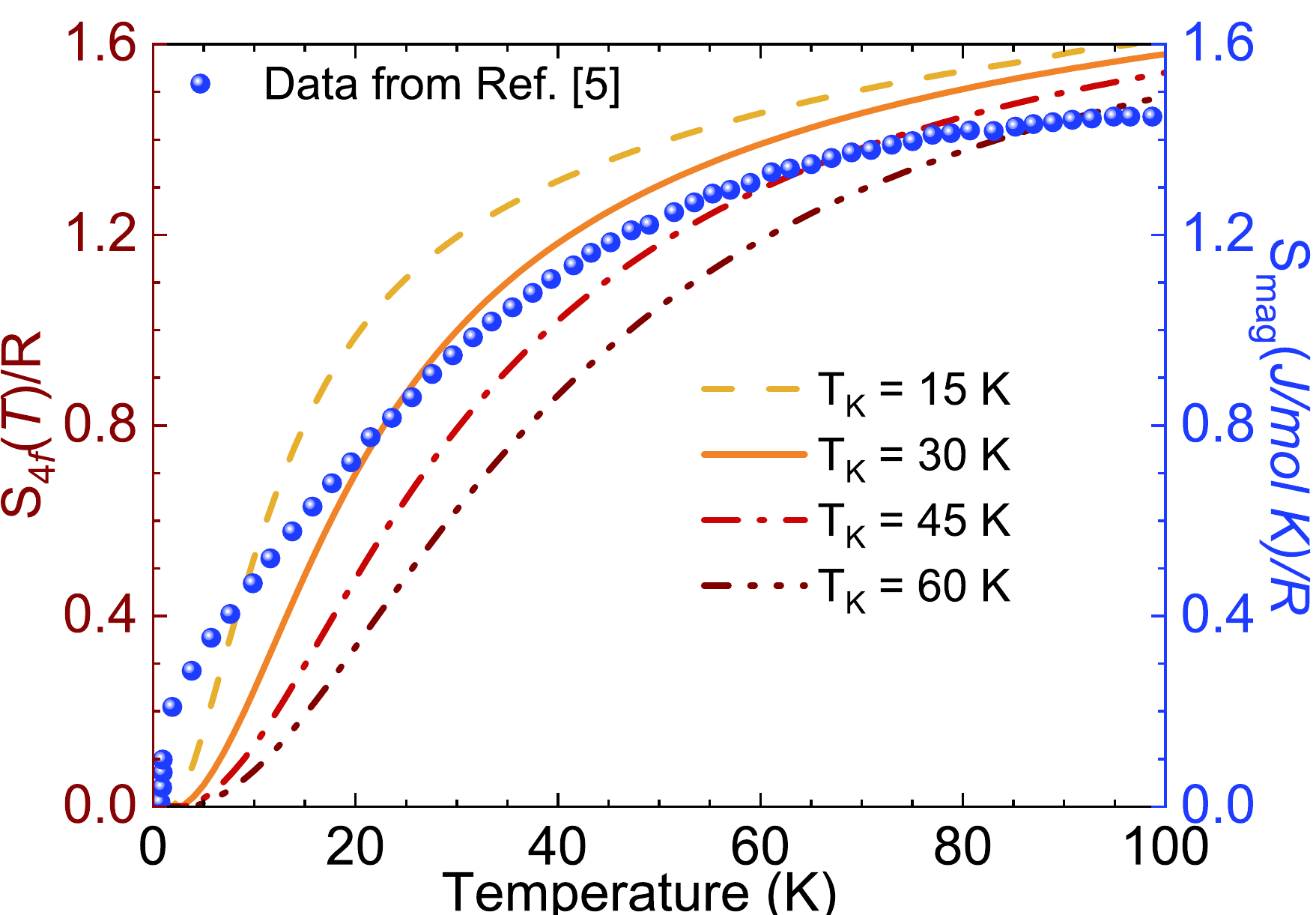}
		\caption{Entropy calculated with SIAM/NCA for different values of $T_K$ compared to the data from entropy from Ref.\,\cite{Khim2021}.} 
		\label{fig7}
	\end{figure}
	
	The contribution of excited crystal-field states to the orbital mixed ground state depends on the Kondo temperature. With increasing $T_K$, this contribution rises (see Fig.\,\ref{fig6}\,(a)-(c)). For $T_K$\,=\,15\,K, the calculated linear dichroism (LD) in the 3\,K XAS spectrum still has the wrong sign (see Fig.\,\ref{fig6}\,(d)), thus giving a lower limit for $T_K$. For $T_K$\,=\,60\,K the overall $T$-dependence is similar to the one of $T_K$\,=\,30\,K, but there is a general decrease of the LD signal (see Fig.\,\ref{fig6}\,(f)) which can be expected since in the limit that $k_B$$T_K$ is much larger than the crystal field splitting the LD will eventually become zero with no $T$-dependence, thereby reflecting the spherically symmetric sixfold degenerate Hund's rule ground state. Fig.\,\ref{fig7} shows the calculated entropy within the SIAM/NCA formalism for different values of $T_K$.

	\subsection{NIXS}
\begin{figure}[h]
		\includegraphics[width=0.95\columnwidth]{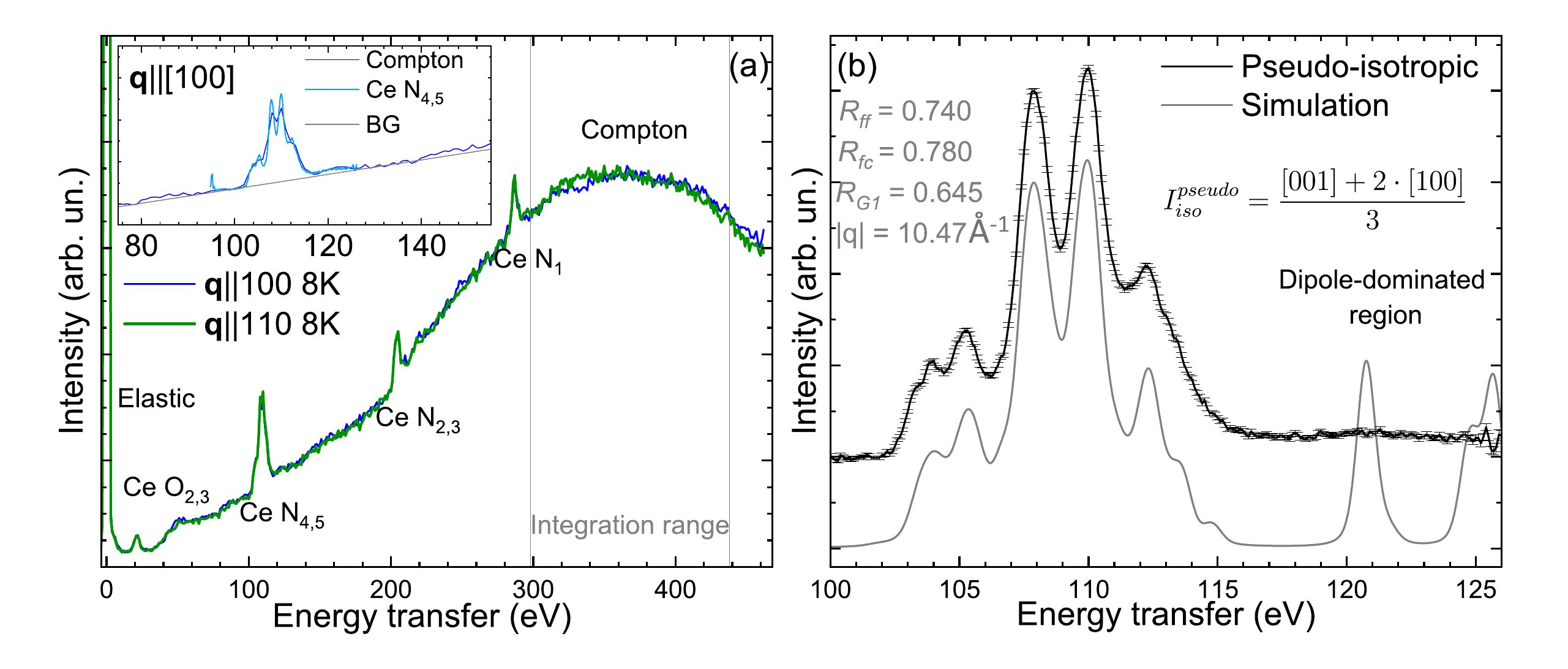}
		\caption{(a) Experimental NIXS spectra of CeRh$_2$As$_2$, displayed in a wider energy window, showing the Compton profile used for normalization and Ce N-edges on top. Linear background correction of the $N_{4,5}$ edge using the wide spectrum, see inset. (b) (Pseudo)isotropic spectrum (black) as constructed from orientation dependent data and optimized simulation (grey). \textit{R$_{ij}$} refer to reduction factors and $\mid q \mid$ to the momentum transfer in the simulation. }
		\label{fig8}
	\end{figure}

	Fig.\,\ref{fig8}\,(a) shows the NIXS data of CeRh$_2$As$_2$ in a wide energy range showing the strong elastic signal at zero energy transfer, the Compton signal and the $n=4$ intra-shell core hole excitations, namely $N_{4,5}$ ($d \rightarrow f$), $N_{2,3}$ ($p \rightarrow f$) and $N_1$ ($s \rightarrow f$), the most intense and structured being the $N_{4,5}$. To normalize and average different scans, we fix an integration range over the Compton scattering and normalize to its area. Further, to treat the inclined background beneath the edge of interest, we overlay the detailed edge scan to a wide Compton scan measured immediately before or after. We then subtract a linear background considering the pre-edge and a reasonably far energy transfer above edge to avoid capturing post-edge effects (generally related to the detailed electronic structure of the material). The energy alignment process is described in detail in Ref.\,\cite{sundermann2019f}.
	
	Fig.\,\ref{fig8}\,(b) shows the constructed (pseudo)isotropic NIXS spectra at the Ce $N_{4,5}$-edge and the simulation resulting from optimizing the reduction factors of the Slater integrals and the magnitude of $\vec{q}$. We increase $\mid\vec{q}\mid$ to account for the extended radial wave function in the solid with respect to the atomic model. In particular, the reduction of the $G^1$ Slater integral tunes the position of the dipolar contribution, so we select its value based on aligning the (under-broadened) dipole excitation to the broad bump around 122\,eV. In total the reduction factors agree well with previous analysis\,\cite{Sundermann2019}.
	
\begin{figure}[]
		\includegraphics[width=0.6\columnwidth]{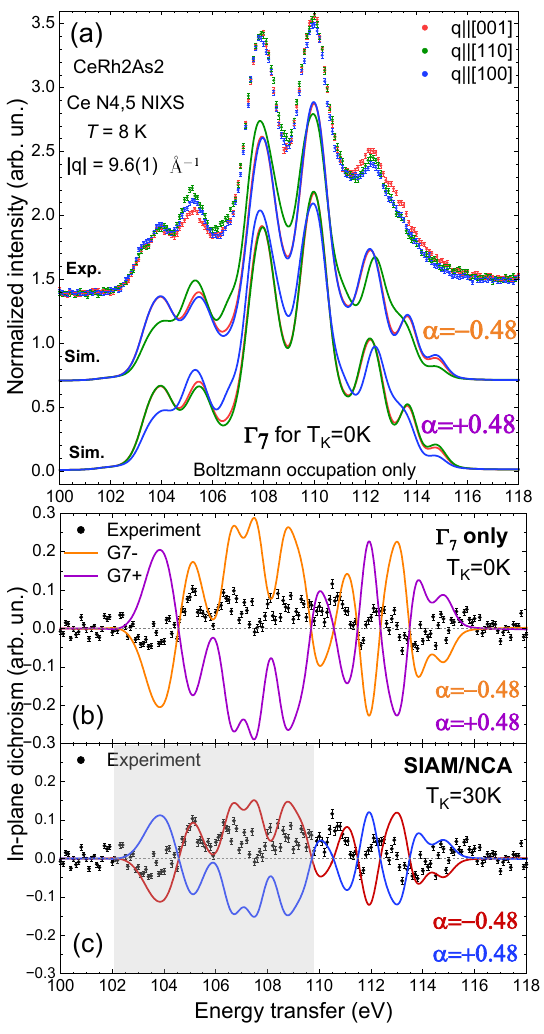}
		\caption{(a) NIXS directional dependent data with simulations for pure $|\Gamma_7\rangle$ states for $T_K$\,=\,0\,K for $\alpha$\,$<$\,0 and $>$\,0, and $|\alpha$\,=\,0.48 giving the amount of $|J_z=\pm5/2\rangle$ in the ground state. (b)\&(c) Experimental $ab$-plane dichroism $I_{001}$\,-\,$I_{100}$ (black dots). Simulation for $|\Gamma_7\rangle$ states with $\alpha$\,$<$\,0 and $>$\,0 as in (b) and in the presence of the Kondo effect according to SIAM/NCA calculations in (c).}
		\label{fig9}
	\end{figure}

	Fig.\,\ref{fig9}\,(a) exhibits the experimental NIXS data and the simulation for $T_K$\,=\,0\,K. We show the simulation for the two possible pure $|\Gamma_7\rangle$ ground states that have the same $\alpha^2$, i.e. the same charge densities, but different signs of $\alpha$. The sign refers to a rotation by 45$^{\circ}$ whereby $\alpha$\,$<$\,0 stands for lobes along [110] and refers to $|\Gamma_7^-\rangle$. Panel (b) and (c) show the experimental and calculated in-plane difference spectra. The simulation of the difference spectra for the pure $|\Gamma_7\rangle$ states without considering the Kondo effect clearly overestimate the directional dependence but the dichroism of the state with $\alpha$\,$<$\,0  ($|\Gamma_7^-\rangle$) has the correct sign in the pre-edge [orange lines in (b)]. The contribution of the rotational invariant $|\Gamma_6\rangle$ state and of the highest crystal-field state, the $|\Gamma_7^+\rangle$ state at 180\,K (15.5\,meV) with opposite dichroism, reduces the \textit{ab}-plane anisotropy so that the simulation for the Kondo mixed state with $\alpha$\,$<$\,0 ($|\Gamma_7^-\rangle$ state) as lowest state fits quantitatively quite well in the pre-edge [red lines in (c)] whereas the assumption of $\alpha$\,$>$\,0 yields a dichroism with the wrong sign. Also in the interval of 106 to 109eV the experimental dichroism is positive as is the simulation based on $|\Gamma_7^-\rangle$. For larger energy transfers, the noise due to the increased Compton background (which has been subtracted) hampers a reliable measurement of the LD. 
	
	\newpage
	\subsection{Band structure}
	A more detailed description of the simple approximation scheme for the Anderson impurity Hamiltonian\,\cite{Zevin1988,Zwicknagl1990} can be found in Ref.\,\cite{Amorese2020} where it was applied to CeCu$_2$Si$_2$. It estimates the contributions of the higher lying crystal-field states into ground state.  
	
\begin{figure*}[h]
		\includegraphics[width=0.68\columnwidth]{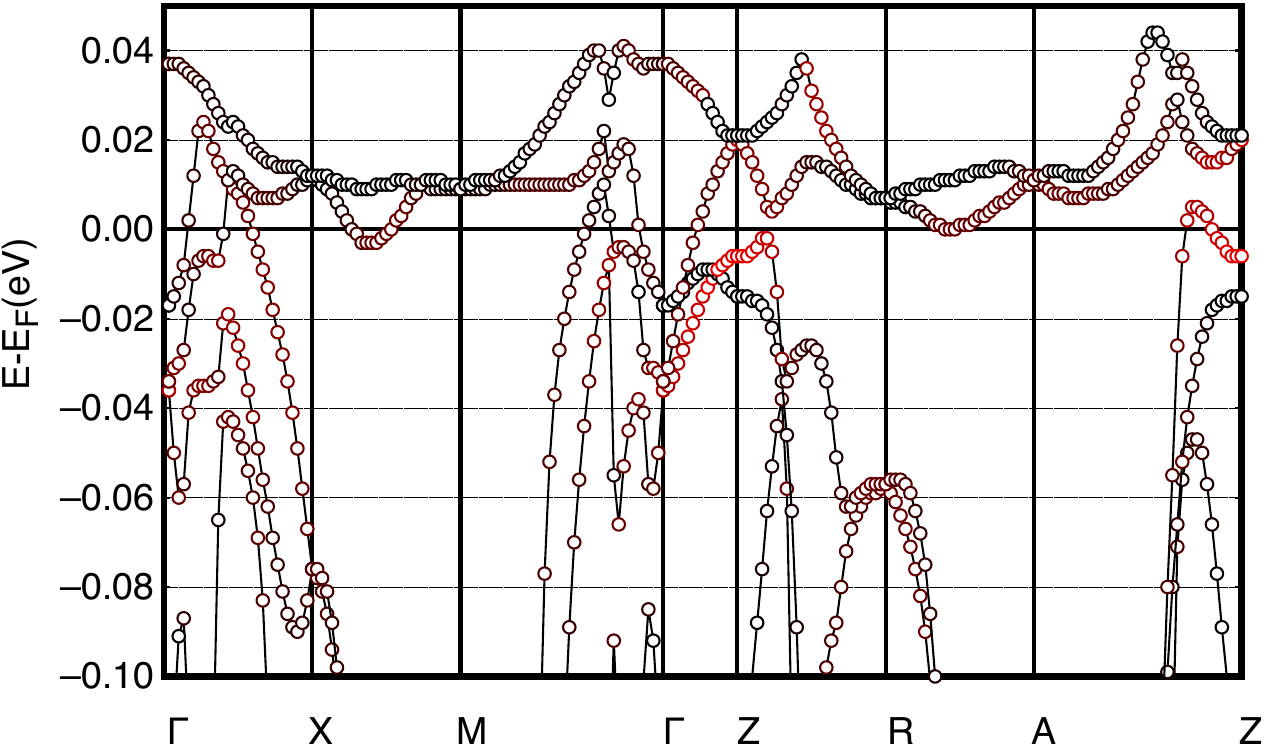}
		\includegraphics[width=0.6\columnwidth]{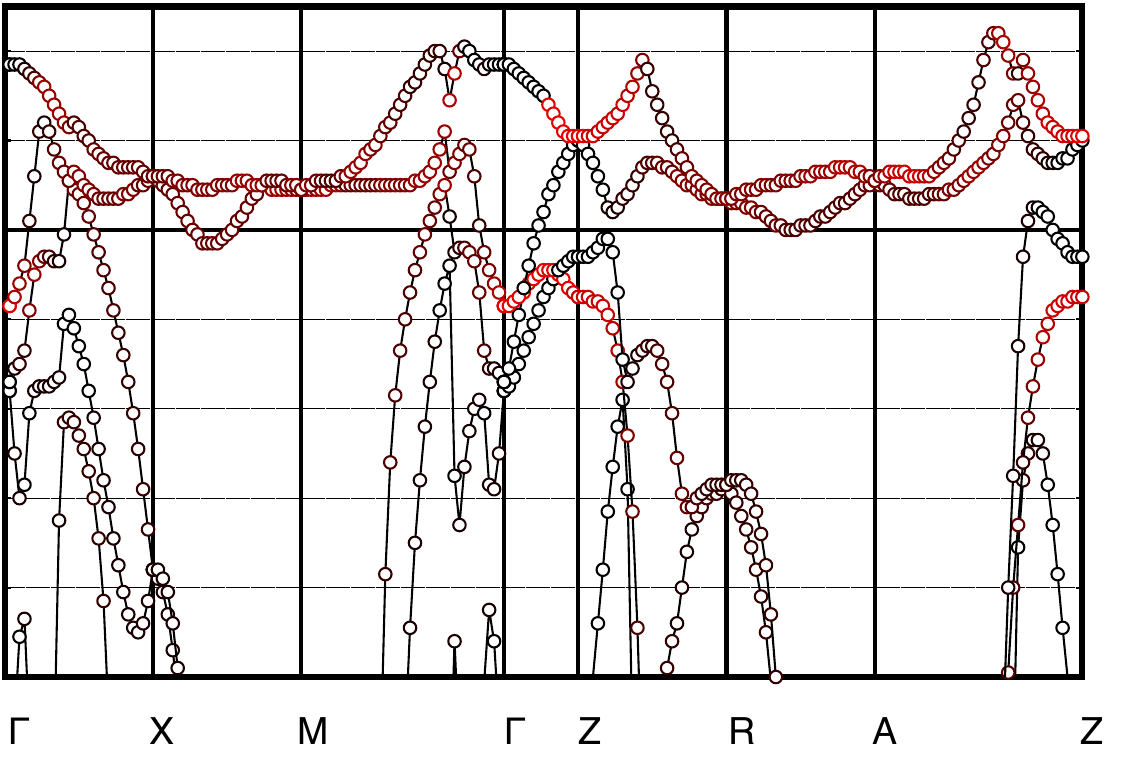}
		\includegraphics[width=0.68\columnwidth]{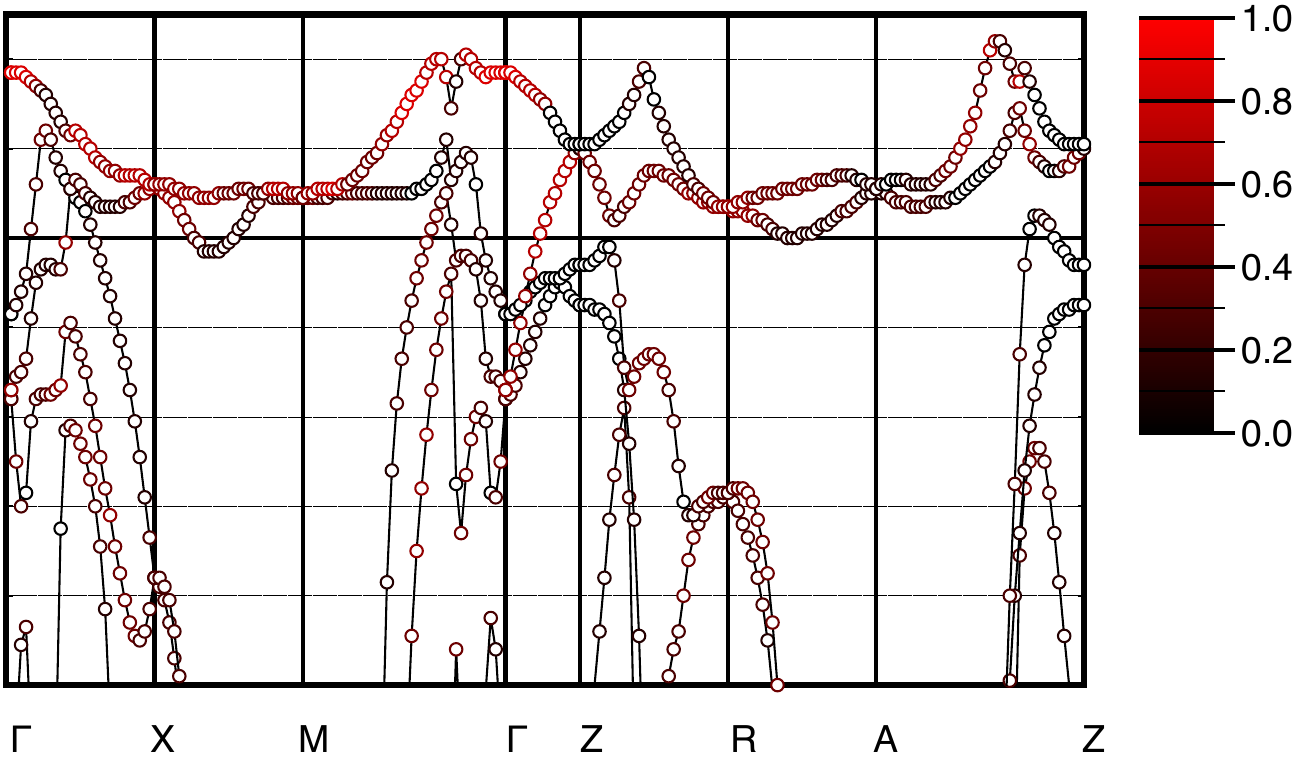}
		\caption{(left)  Renormalized quasiparticle bands, showing the contributions of each crystal-field state; left for $|\Gamma_7^-\rangle$, middle for $|\Gamma_6\rangle$ and right for $|\Gamma_7^+\rangle$, red standing for high and black for low occupation.}
		\label{fig10}
	\end{figure*}

	Also the renormalized band structure calculations\,\cite{Zwicknagl1992} in Fig.\,\ref{fig10} support the hybridization induced multi-orbital ground state. Here the contributions of the three crystal field states (ground state and excited 4$f$ states at 30 and 180\,K) are projected out to the respective bands. It turns out that the heavy band at zero energy with mainly $|\Gamma_7^-\rangle$ character has also non-negligible contributions of $|\Gamma_6\rangle$ and $|\Gamma_7^{+}\rangle$ states, see Fig.\,\ref{fig10}.


\providecommand{\noopsort}[1]{}\providecommand{\singleletter}[1]{#1}%
%


\end{document}